\documentclass[fleqn,usenatbib]{mnras}

\usepackage{newtxtext,newtxmath}

\usepackage[T1]{fontenc}

\DeclareRobustCommand{\VAN}[3]{#2}
\let\VANthebibliography\thebibliography
\def\thebibliography{\DeclareRobustCommand{\VAN}[3]{##3}\VANthebibliography}

\usepackage{graphicx}	
\usepackage{amsmath}	

\def\msun{\hbox{M$_\odot$}}
\def\tdyn{\hbox{$\tau$$_{\rm dyn}$}}

\title[Star Cluster Ecology]{Star Cluster Ecology: Revisiting the Origin of Iron and Age Complex Clusters}

\author[N. Bastian \& J. Pfeffer]{
Nate Bastian$^{1,2,3}$\thanks{E-mail: nate.bastian@dipc.org} and
Joel Pfeffer$^{4}$
\\
$^{1}$ Donostia International Physics Center (DIPC), Paseo Manuel de Lardizabal, 4, 20018, Donostia-San Sebasti\'an, Guipuzkoa, Spain\\
$^{2}$ IKERBASQUE, Basque Foundation for Science, 48013, Bilbao, Spain \\
$^{3}$Astrophysics Research Institute, Liverpool John Moores University, IC2 Liverpool Science Park, 146 Brownlow Hill, Liverpool L3 5RF, UK\\
$^{4}$ICRAR M468 The University of Western Australia 35 Stirling Hwy, Crawley Western Australia 6009, Australia\\
}

\date{Accepted XXX. Received YYY; in original form ZZZ}

\pubyear{2021}

\begin{document}
\label{firstpage}
\pagerange{\pageref{firstpage}--\pageref{lastpage}}
\maketitle

\begin{abstract}
Typical globular clusters (GCs - young and old) host stellar populations with little or no star-to-star variations in heavy elements (e.g., Ca, Fe) nor in age.  Nuclear star clusters (NSCs), on the other hand, host complex stellar populations that show multi-modal distributions in Fe and often in age, presumably due to their unique location at the centre of a large galactic potential well.  However, recently a new class of clusters have been discovered, exemplified by the clusters Terzan~5 and Liller~1, two high mass, high metallicity clusters in the inner Galactic regions.  It has been suggested that these are not true GCs, but rather represent left over fragments of the formation of the Galactic Bulge.  Here, we critically assess this scenario and find that the role of dynamical friction likely makes it untenable and that the method used to estimate the initial masses of the clumps was invalid.  Instead, it appears more likely that these clusters represent a relatively rare phenomenon of existing GCs accreting gas and forming a 2nd generation, as has been previously suggested. 


\end{abstract}




\section{Introduction}

Over the past twenty-odd years, there has been a unification of stellar cluster types, which were previously thought to be unique and separate.  Hence, while there appears to be a diverse ecology of clusters, advances in theory and simulations have emphasised their connectedness, representing different parts of a larger continuous parameter space.  Examples include linking the young massive clusters (YMCs - \citealt[][]{pz10}) with their lower mass counterparts, the Galactic open clusters (OCs), through a continuous mass distribution.  Likewise, the Galactic open clusters represent the young (and typically lower mass) end of a continuous age distribution that naturally extends into the classic globular cluster population (once cluster disruption is taken into account - \citealt{elmegreen97}).  This can be seen most clearly in nearby galaxies, such as the Large and Small Magellanic Clouds, whose cluster populations suggest a continuum of formation throughout cosmic time, with cluster formation tracing the overall chemical build up of the galaxies over time \citep[e.g.,][]{horta21}.

Numerical simulations have also had recent success in linking YMCs/OC to GCs \citep[e.g.,][]{li17, pfeffer19, choksi19, lahen20}, explaining both sets of objects with the same formation/evolution physics \citep[][]{kruijssen15}.  Whether or not a massive cluster formed is not directly linked to cosmic age, rather it is the local conditions that determine the mass of clusters that form.

Beyond this framework, there is another class of cluster, the nuclear star clusters \citep[NSCs - e.g.,][]{neumayer20}.  These clusters, typically residing (or have resided) at the centre of their host galaxy, are some of the most massive and dense clusters known.  They also have complex formation histories, which appear to be linked to the properties of their host galaxies.  For massive host galaxies, their NSCs appear to be formed primarily in-situ, i.e., multiple epochs of star-formation happen within the cluster.  For lower mass galaxies, their NSCs are dominated by GC in-spiralling and mergers \citep{turner12, neumayer20}.  There are a number of clusters in the MW (beyond the NSC at the centre of the Galaxy), which have been captured during the accretion event of a nucleated dwarf galaxy \citep[][]{pfeffer21}.

Finally, there are two clusters which do not appear to fit in well with the taxonomy described above.  The clusters in question are Terzan~5 and Liller~1, both of which host significant iron spreads ($\sim1$~dex)\footnote{Although in the case of Liller~1 further spectroscopic confirmation of the Fe-spreads is required.} and age spreads ($\sim5-8$~Gyr) within their stellar populations \citep[][]{ferraro16, ferraro21}.  This could imply that they may be accreted NSCs.  However, their current orbits are consistent with an in-situ formation channel, i.e., they do not appear to be part of an accretion event.  Additionally, their highest metallicity components ([Fe/H]$\sim$ solar - \citealt[][]{origlia11}) also suggest that they were not formed within a dwarf galaxy, as these galaxies typically do not reach these high metallicities until the current epoch (i.e., they should have been very low mass galaxies $\sim12$~Gyr ago when their first stellar populations formed - e.g., \citealt[][]{kruijssen19}).

As a potential explanation for these two anomalous clusters, \citet{ferraro09} and \cite{ferraro21} have suggested a new formation channel, namely that these two clusters are remnants of large, gravitationally bound, long lived clumps, most of which eventually merged to form the bulge of the Milky Way.  In this sense, these clusters would represent pristine fragments of the galactic bulge, which can be used to trace the formation and evolution of the bulge.

In the present work we take a critical look at the "pristine fragments" theory as well as alternative theories which have been put forward in the literature.  This paper is organised as follows: in \S~\ref{sec:PFGB} we outline the basic scenario of the "Pristine Fragments of the Galactic Bulge" theory while in \S~\ref{sec:problems} we discuss problems with the suggested theory.  In \S~\ref{sec:alternative} we briefly discuss an alternative theory that has been put forward in the literature and in \S~\ref{sec:conclusions} we summarise our findings.

\section{Theory of Pristine Fragments of the Galactic Bulge (PFGB)}
\label{sec:PFGB}

Observations of high-z spiral galaxies often show bright UV clumps within them \citep[e.g.,][]{cowie95, giavalisco96, vanDenBergh96, elmegreen04, elmegreen05, guo15, Shibuya_et_al_16}, indicative of massive star-forming complexes.  It is thought that many of these clumps will migrate to the centre of the host galaxy, due to dynamical friction, and deposit their stars there, effectively building the bulge of these proto-galaxies \citep[e.g.,][]{Immeli04, Dekel09}.  In fact, it has been suggested that many of the metal rich globular clusters may form in such clumps \citep[e.g.,][]{shapiro10}.

\citet{ferraro09} and \citet{ferraro21} have suggested that these clumps may be long lived ($>5-8$~Gyr), implying that they are gravitationally bound and separate structures within the host galaxies.  The authors adopt a mass-metallicity relation from galaxy surveys, and use this to estimate that the initial masses of these clumps were $\gtrsim10^9$~\msun.  In this model, the massive clusters Terzan~5 and Liller~1 are the remnants of these initial high mass clumps.  Due to the high masses of the clumps, the authors suggest that they would be able to retain stellar ejecta and SNe and form new generations of stars which would largely mirror the formation and chemical enrichment history of the bulge.

In a sense, in this scenario Terzan~5 and Liller~1 could be seen as the in-situ nuclear star clusters of these clumps.  However, the authors suggest that their formation mechanism is unique, hence they may not host light element abundance spreads within them (a.k.a., multiple populations).  However, as the origin of multiple populations is still unknown  \citep[e.g.,][]{bl18}, it is unclear if the remnant clusters from such a clump would be expected to host such populations or not.  In the case of Terzan~5, an initial study by \citet[][]{origlia11} did not find evidence for multiple populations within the cluster using the Al-O anti-correlation as a diagnostic.  However, \citet[][]{pancino17} found that Al did not vary significantly in metal rich clusters, meaning that this is not the best element to use when looking for multiple populations in clusters like Terzan~5.  On the other hand,  \citet[][]{schiavon17} and \citet[][]{nataf19} found the normal CNO-based (anti-)correlations typical of globular clusters within the older stellar population of Terzan~5, suggesting that it a typical GC (within that population).

One potential caveat, however, is that recent results have shown, that younger clusters appear to host smaller light element abundance spreads (and below $\sim2$~Gyr no spreads have been detected), hence it is possible that only the older stellar population within the clusters would be expected to show (large) light-element abundance spreads \citep[][]{martocchia18}.  For Terzan~5, the younger population (at $\sim4.5$~Gyr) should be old enough to show signatures of light-element abundance spreads, while for Liller~1, it is possible that the younger population is too young to show such signs.

\section{Problems with the PFGB Theory}
\label{sec:problems}

\subsection{The estimate of the host clump mass}

In order to estimate the initial mass of the clump, \cite{ferraro09} and \cite{ferraro21} used the metallicity of the main stellar population in each cluster and apply the (redshift dependent) mass-metallicity relation of galaxies \citep{Mannucci09}.  Doing this, the authors arrive at an initial mass of $\sim3\times10^9$~\msun.

However, the adopted mass-metallicity relation is for full galaxies, not individual regions within a galaxy, i.e., it only applies to self-enriching systems, not disc instabilities.  The initial metallicity of the clump would not be set by its own mass, but rather simply from the metallicity of the gas clouds within the host galaxy from which it formed.  The metallicity of this gas in turn was set by the chemical enrichment of the Milky Way up until that time.  Hence, by adopting the metallicity of the main stellar population, what the authors are actually estimating is the mass of the Milky Way (assuming the clusters formed within the Galaxy) at the time of formation, not an individual clump within it.

Hence, based on the metallicity of the cluster/clump, it is not possible to estimate its initial mass based on the mass-metallicity relation of galaxies.  In the same way, the metallicity of globular clusters is not linked to their mass, rather it is the metallicity of the gas within the host galaxy, at the time of their formation.

Without such high initial masses of the clumps, most aspects of the PFGB theory would not operate.  


\subsection{Chemical evolution}

\citet{ferraro21} suggest that the PRGBs will be able to retain the stellar and SNe ejecta of their constituent stars (along with accretion from the surroundings) hence they will chemically evolve along an evolutionary track, in a similar way as dwarf galaxies.  As discussed above, \citet{ferraro21} estimate a stellar mass of $\sim10^9$~\msun\ for a typical clump.  The authors argue that dwarf galaxies of similar stellar mass are able to chemically enrich themselves, hence the clumps would be expected to do the same.

However, galaxies with stellar masses of $10^9$~\msun\ will typically have large dark matter halos around them, with a total mass of $\sim10^{11}$~\msun \citep[e.g.][]{Behroozi19}.  It is this dark matter halo that allows the galaxy to retain SNe ejecta and chemically evolve accordingly. Bulge fragments, on the other hand, would not possess such a halo, hence we would not expect them to enrich significantly through SNe within their own populations.  Hence, without a large dark matter halo surrounding the clump in order to retain stellar ejecta, the PFGB theory would not be able to explain the chemical evolution of the stellar populations within clusters like Terzan~5 and Liller~1 \footnote{This is in contrast with the NSCs which were the central clusters of dwarf galaxies which did contain large dark matter halos, allowing for such self-enrichment.}.

As an example, a number of young and intermediate age stellar clusters are known with masses between $10^7 - 10^8$~\msun.  These include W3 in NGC~7252 \citep[][]{maraston04}, C1 in NGC~34 \citep[][]{schweizer07}, and c113 in NGC~1316 \citep[][]{bastian06}.  None of these clusters show evidence of retaining gas/dust, either through their light profiles \citep[][]{longmore15}, nor do they show evidence of extended star-formation histories \citep[][]{cz14, cz16}.  Hence, objects with these stellar masses, that do not host large dark halos, are unable to retain/accrete significant amounts of gas within them \footnote{While many models have been put forward that invoke multiple generations of star-formation within typical GCs to explain the origin of multiple populations, observations of young massive clusters as well as the properties of full populations of ancient GCs do not support such scenarios \citep[see the recent review in][]{bl18}. }.

\subsection{Dynamical Friction}

While we have argued above that the estimate of the initial clump masses of $\sim10^{9}$~\msun\ are invalid, here we adopt these values to see, if they were true, whether such clumps could survive long enough to be observable today.

A critical question is whether large, gravitationally bound fragments in the early life of the Galaxy could survive for long enough ($\geq 8$~Gyr) to be able to retain/accrete gas and form new generations of stars. The key timescale for this to be valid is the dynamical friction timescale, \tdyn. If \tdyn\ is too short, the clump will spiral into the Galactic centre and dissolve.  Hence, \tdyn\ must be longer than, or approximately equal to the age of the system if it is to survive to the present day.

In order to estimate \tdyn\ for the initial clumps we followed \citet[][]{lc93} (their Appendix~B), adopting an isothermal sphere as the Milky Way potential with a circular velocity V$_{\rm c} \approx 230$~km/s.  For the fragments we adopted masses of $10^9$~\msun\ (a lower limit to the estimates of \citealt{ferraro21}), a circular orbit and initial galactocentric radii between 3 and 11~kpc.  Under these assumptions \tdyn\ is 0.15, 0.67, and 1.5~Gyr for initial radii of 3, 7 and 11~kpc, respectively.  Hence, for a fragment of this mass, dynamical friction would act extremely rapidly, meaning that it would spiral into the centre of the Galaxy within a fraction of the time needed for it to survive.  If the clump formed well outside 11~kpc it could have survived to the present day, however it would not be expected that such clumps could form at such large galactocentric distances.  Hence, the formation and long term survival of such a clump can be ruled out at high significance.

Even for an initial mass of $10^8$~\msun, \tdyn$=0.9,4.4,$ \& $10.2$~Gyr, for initial radii of 3, 7 and 11~kpc, respectively.  If the clump needed to survive for $>8$~Gyr, then it must have began at extreme galactocentric radii ($\gtrsim10$~kpc), otherwise it would be expected spiral into the centre of the Galaxy too rapidly to account for the observed characteristics.  Hence, even if the fragment was undergoing severe mass-loss (resulting in the loss of $\sim90$\% of its initial mass), it would still rapidly sink to the center of the Milky Way.

Of course there is a sweet spot, where if the clumps underwent heavy mass loss they could avoid spiralling in to the centre of the Galaxy, i.e., that mass loss and dynamical friction happen in such a way to allow the clump remnants to survive, but to be brought into the inner regions of the Galaxy.  However, we note that in the PFGB scenario the clump must keep its large mass for an extended period of time (likely at least a few Gyr) in order to retain gas within it, to form a second generation of stars $5-8$~Gyr later.  This places further constraints on how quickly it could lose mass in order to avoid the destructive effects of dynamical friction.

On the other hand, by adopting an initial mass of $2\times10^6$~\msun, i.e., similar to the current masses of Terzan~5 and Liller~1, we find that dynamical friction timescale is longer than the Hubble time already for initial galactocentric radii of $\sim2$~kpc.  Hence, if the clusters formed with approximately the current mass, we would expect them to be long lived objects.

\subsection{High redshift observations}

Beyond the arguments outlined above, there has been a number of studies that have investigated whether the UV-clumps observed at high-redshift are long lived objects or whether they are rapidly destroyed, either through dynamical friction or dissolution into the host galaxy (potentially driven by stellar feedback within the clumps).  If the clumps were long lived we would expect to see them not just in the UV (indicative of young ages) but also in the optical as they age.  In fact, we may expect to see a number of these remnant clumps today in nearby galaxies.

One of the most systematic studies was performed by \citet{Shibuya_et_al_16}, who used HST imaging to investigate the evolution of clumpy galaxies from $z=0$-$10$. 
They find that the redshift evolution of the fraction of clumpy galaxies in the rest-frame UV broadly traces the cosmic star formation rate density. This is consistent with the scenario where the clumps form via disc instabilities in star-forming galaxies. 

At $z=1$-$3$ the fraction of galaxies that have clumps in the rest-frame UV is 50-60 per cent (with a lower fraction in `quiescent' galaxies with lower SFRs). However, only $\approx10$ per cent of galaxies have clumps in rest-frame optical over the same redshifts, indicating that the clumps are not long lived.
The fraction of galaxies with optical clumps could instead be explained by galaxy mergers (see their Figure~5; see also \citealt{Zanella19}).

\citet{Shibuya_et_al_16} also investigated the $m_\mathrm{UV} - m_\mathrm{opt}$ colour (an indication of age) of clumps as a function of galactocentric radius. Only very massive galaxies ($>10^{11}$~\msun) show a correlation between clump colour and radius, such that clumps are redder at smaller distances. They found no strong correlation in galaxies with stellar masses $10^9$-$10^{11}$~\msun, indicating that the clumps may be rapidly destroyed before significant clump migration.
This is in agreement with the lifetimes estimated for high redshift clumps, which are probably not more than $\sim$500~Myr on average due to a combination of feedback, dynamical friction and tidal disruption \citep[e.g.][]{Genzel11, Zanella19}.

We conclude that the UV clumps are likely not long-lived objects and would not be able to survive the $\sim5-8$~Gyrs needed in order to form the observed second generation of stars within them, i.e., as seen in Terzan~5 and Liller~1.


\section{Alternative}
\label{sec:alternative}

As an alternative explanation for clusters like Terzan~5 and Liller~1, \citet[][]{mb18} have suggested that massive clusters in the inner galaxy may have orbits which rarely align (within some tolerance) in position and velocity with giant molecular clouds (GMCs).  Once such a chance super-position happens, the massive cluster may be able to accrete large amounts of gas/dust rapidly from the GMC.   If this accretion is rapid and dense enough, it can overcome the stellar feedback from the original stellar population, and may be able to form a second (or more) generation of stars.  This is largely analogous to the in-situ formation within nuclear star clusters observed today.  

Running a series of simulations, \citet[][]{mb18} found that such an alignment may happen once or twice per Hubble time, with the likelihood a strong function of cluster mass (with more massive clusters having a higher chance of gas accretion).  Hence, this is an effect that would be expected to happen predominantly for massive clusters orbiting within the inner few kpc of the Galaxy, where the gas density is highest.  Additionally, the probability would be highest for GCs with disk-like orbits (as the GMCs are expected to be largely confined to the disk) as found for both Terzan~5 and Liller~1 \citep{Massari15, Baumgardt19} \footnote{While \citet{massari19} classify Terzan~5 as a bulge GC, this is due to it being located in the inner 3.5~kpc of the Galaxy, irrespective of its orbit.}.

Typically, other than nuclear star clusters \citep[][]{neumayer20}, clusters (open clusters, YMCs or GCs) are not able to accrete significant amounts of gas from their surroundings.  This is due to a combination of ram pressure and stellar feedback (winds, SNe, and photionisation), which act to expel the ejecta from stars within the cluster and stop accretion from the surroundings \citep[e.g.,][]{chantereau20}.  Additionally, most GCs have orbits away from the disk, both spatially and kinematically.  However, as seen in nuclear star clusters, if a large amount of gas can be dumped onto the cluster in a short time, this can overwhelm the stellar feedback, allowing the gas to avoid being photoionised and overcome the outflowing stellar feedback.  Due to the relative velocities of the NSC and the gas cloud, ram pressure stripping would also not act in the  \citet[][]{mb18} scenario.  In such a case, the gas may cool (if it is optically thick) and form a new generation of stars.  Depending on how much gas is dumped onto the original cluster, the second generation of stars could make up anywhere between a small and large fraction of the final cluster stellar mass. {One caveat to this scenario is that such a close passage to a GMC may actually disrupt the cluster before a large amount of gas could be transferred onto it.  However, it would be the most massive GCs that would have the highest chance of surviving such an event \citep[e.g.][]{Gieles_06, Elmegreen_and_Hunter_10}.}

This scenario would suggest that the older populations present within Terzan~5 and Liller~1 would be the remnants of classic globular clusters (likely of high mass) and the younger populations would be formed from gas accreted by the ancient cluster due to the GMC interaction.  We note that there is expected to have been a population of GMCs in the disk of the Milky Way, continuously for the past $7$~Gyr or more \citep[given the relatively constant star-formation rate of the Galaxy over that period, e.g.][]{Snaith15}.  The GMCs are not expected to have bulge like kinematics/orbits, rather they would be part of a disk.  The interaction and subsequent accretion from a GMC onto an existing GC would then only happen when the GCs orbit passed through the disk and happened to have a very low relative velocity.  In fact, both Terzan~5 and Liller~1 have orbits confined to the plane of the Galactic disk \citep[typically only reaching $\sim 200$~pc above the disc plane,][]{Massari15, Baumgardt19}, which may increase the likelihood of such a chance encounter for these GCs.  {The larger the vertical scale height, or the further the radial apocentre/pericentre of the GC orbit, would lead to a reduction of the probability of a GC/GMC interaction that could lead to the accretion of gas by the GC, due to the centrally peaked and thin disk of GMCs in the Milky Way \citep[e.g.,][]{ke12}. }
This may explain why the other very massive ($M> 10^6$~\msun) bulge GCs NGC 6388 and NGC 6441 do not similarly show multiple iron peaks. Both GCs have orbits which are inclined to the Galactic disc by $\sim30$ degrees \citep{Baumgardt19}, reaching $\approx 1$~kpc above the disc plane, which would significantly lower their probabilities of GC/GMC interactions.

In passing, we note that star formation is likely to have been happening within this gas-rich (GMC-rich) disk throughout the history of the Galaxy, however these stars have not been perturbed onto orbits significantly out of the disk due to the Milky Way's relatively quiescent merger history over the past $>7$~Gyr \citep[e.g.][]{Wyse01, Hammer07, Stewart08, Kruijssen20}.

A prediction of such a scenario is that it should be relatively rare, i.e., even for massive clusters within the inner Galaxy, we would only expect a handful of such clusters to have undergone such a process.  Additionally, as these would be rare events, we would expect to see largely discrete star-formation events.  If the star-formation histories of Terzan~5 and Liller~1 were to be found to be extended, or continuous over many Gyr, this would strongly disfavour the \citet[][]{mb18} scenario.  One caveat to this point, however, is that while in most clusters the capture rate of field stars by a GC is low \citep[e.g.,][]{mieske07}, for high mass clusters in the inner regions of the Galaxy this can begin to become significant, due to the much higher field star densities.  Hence, for clusters like Terzan~5 and Liller~1 we may expect a background population of stars, over a range of metallicities (and possibly ages), to be present within the clusters, although these populations should be sub-dominant.  This could explain the purported third, minor, low metallicity population \citep[e.g.,][]{massari14}.  Also see \citet{Khoperskov18} for a discussion of the transfer of stars between GCs on disc orbits.

\section{Conclusions}
\label{sec:conclusions}

We have critically assessed theories for the origin of the iron/age complex stellar clusters, Terzan~5 and Liller~1 in order to better understand their origin.  \citet{ferraro09} and \citet{ferraro21} have suggested that such clusters are the remnants of fragments of the Galactic bulge, hence they would represent a unique window into the formation of the bulge.  Critically, for this theory to be valid, the initial fragment must have a mass far in excess of the current cluster mass, and it must have been able to survive for at least a Hubble time.  We showed that the method used to estimate the initial clump masses was invalid, as the adopted mass-metallicity relation of galaxies would not correspond to a specific sub-region (clump) with the galaxy, but rather to the full galaxy.  In this way, the estimate mass of the clumps would correspond to the stellar mass of the Milky Way at the time of formation, not the mass of an individual clump.

The critical timescale is that of dynamical friction, which is the timescale that any such clump would spiral into the Galactic centre.  For the estimated initial mass of the fragments ($\sim10^9$~\msun; \citealt[][]{ferraro21}) we estimate dynamical friction timescales of $<1.5$~Gyr for initial galactocentric radii of 11~kpc or less.  This is far shorter than the estimated duration of star-formation (hence survival) of the fragments, even when assuming severe mass loss during its evolution.  This effectively rules out this class of theory as the origin of these anomalous clusters.

Instead, we favour the theory of \citet[][]{mb18} where, under certain, relatively rare conditions, a globular cluster is able to accrete a significant amount of gas/dust from its surroundings (essentially when the cluster and GMC orbits coincide in time, velocity and space) and form a second generation of stars.  Since the gas accreted would be from the Galaxy, it would have undergone the same chemical enrichment processes and follow the same track as the rest of the inner Galaxy.  Hence, we would expect the 1st and 2nd generations of stars within clusters like Terzan~5 and Liller~1 to follow the same age/metallicity relation.  In this scenario, the ancient populations of these clusters would be the remnants of classical GCs and the younger populations would be the 2nd generation stars formed in these rare events.  This scenario predicts that the process should be most likely for high mass clusters with disk-like orbits in the inner regions of the Galaxy, consistent with what is seen in Terzan~5 and Liller~1.  Additionally, the scenario predicts that the large old, metal poor population should show the canonical light element abundance patterns seen in all massive ancient clusters, while the large young population may or may not show such anomalous chemistry.


\section*{Acknowledgements}
The authors would like to that Sara Saracino, Madeleine McKenzie, and Emanuele Dalessandro for valuable comments on the draft and Mark Gieles for discussions on the project. The authors would like to thank the referee for his/her comments which improved the paper.  NB gratefully acknowledge financial support from the European Research Council (ERC-CoG-646928, Multi-Pop)  and from the Royal Society in the form of a University Research Fellowship.
JP is supported by the Australian government through the Australian Research Council's Discovery Projects funding scheme (DP200102574).

\section*{Data Availability}

The data underlying this article will be shared on reasonable request to the corresponding author.



\bibliographystyle{mnras}
\bibliography{ecology} 








\bsp	
\label{lastpage}
\end{document}